\documentclass[aps,prl,showpacs,twocolumn,floats,psfig,superscriptaddress]{revtex4}
\usepackage{amssymb}
\usepackage{amsbsy}
\usepackage{amsmath}
\usepackage{epsfig}

\begin{document}

\date{\today}

\title {Commensurate lock-in and incommensurate supersolid phases of hardcore bosons on anisotropic triangular lattices}

\author{Sergei V. Isakov}
\affiliation{Department of Physics, University of Toronto,
Toronto, Ontario M5S 1A7, Canada}

\author{Hong-Chi Chien}
\affiliation{Department of physics, National Tsing Hua University, Hsinchu, Taiwan 30013 R.O.C.}

\author{Jian-Jheng Wu}
\affiliation{Electrophysics
Department, National Chiao Tung University, Hsinchu, Taiwan 300 R.O.C.}

\author{Yung-Chung Chen}
\affiliation{Department of Physics, Tunghai University, Taichung, Taiwan 407 R.O.C.}

\author{Chung-Hou Chung}
\affiliation{Electrophysics
Department, National Chiao Tung University, Hsinchu, Taiwan 300 R.O.C.}

\author{Krishnendu Sengupta}
\affiliation{TCMP division, Saha Institute of
Nuclear Physics, 1/AF Bidhannagar, Kolkata-700064, India}

\author{Yong Baek Kim}
\affiliation{Department of Physics, University of Toronto,
Toronto, Ontario M5S 1A7, Canada}


\date{\today}

\begin{abstract}

We investigate the interplay between commensurate lock-in and incommensurate supersolid
phases of the hardcore bosons at half-filling with anisotropic nearest-neighbor hopping
and repulsive interactions on triangular lattice. We use numerical quantum 
and variational Monte Carlo as well as analytical Schwinger boson mean-field analysis
to establish the ground states and phase diagram.
It is shown that, for finite size systems, there exist a series of jumps between different supersolid phases as the anisotropy
parameter is changed. The density ordering wavevectors are
locked to commensurate values and jump between adjacent supersolids. 
In the thermodynamic limit, however, the magnitude of these jumps
vanishes leading to a continuous set of novel incommensurate
supersoild phases.


\end{abstract}


\maketitle

{\it Introduction.}--- Supersolid phases with coexisting long-range
diagonal and off-diagonal orders have long been recognized as
interesting conceivable ground states of superfluid systems in the
presence of mobile vacancies \cite{oldpapers}. Recent experiments have also
reported possible evidence for such a state in $^4 {\rm He}$
\cite{chan1}. Another, relatively new, route to supersolid phases
has recently been explored in the hardcore boson models on
frustrated two-dimensional (2D) lattices \cite{trilat:ss,old1}. 
In such lattices, supersolid phases may arise due to intricate
competition between kinetic and interaction energies, and 
as a result present an excellent playground for discovery of 
possible novel universality classes of quantum phase transitions \cite{balents1,kagome1}.
Further interest in these systems stems from potential realization of these models 
in ultracold atomic systems on optical lattices \cite{ultra1}.

The diagonal (density) orders of the supersolid phases 
discovered so far have been restricted to commensurate orders. For example,
a commensurate supersolid was discovered for the hardcore bosons on
the isotropic triangular lattice \cite{trilat:ss} with the density ordering 
wave vector $\mathbf{Q}_0=(4\pi/3,0)$ along
with non-vanishing superfluid order parameter \cite{trilat:ss}. 
A qualitative understanding of such a supersolid state can be obtained
by considering the deviation from a commensurate boson filling
fraction at which the ground state is a perfect Mott crystal with a
long-range diagonal order. The additional particles or holes
resulting from such a deviation condense to produce a superfluid
while retaining the backbone of the existing Mott solid. This leads
to the coexistence of off-diagonal (superfluid) and diagonal (Mott)
orders. One may expect that incommensurate version of the
supersolid phases may arise when there are more than one competing 
interactions or length scales in the system. If such a phase exists
in lattice models, this may be a much closer analog of the supersolid
phases originally proposed for the continuum.

In this letter, we investigate possible presence of incommensurate 
supersolid phases of the hardcore bosons at half-filling with anisotropic nearest-neighbor
hopping and repulsive interactions on triangular lattices. We establish the
ground states and phase diagram using several different and complementary 
methods, namely numerical quantum and variational Monte Carlo (QMC) 
techniques as well as analytical Schwinger-Boson mean-field theory.
It is found that, for finite size systems, there exist a series of jumps 
between different supersolid phases as a function
of the anisotropy parameter. The (density) ordering wavevectors of
these phases are pinned to commensurate values and jump upon
entering a nearby supersolid phase. The ordering wavevectors assume
every single commensurate values for an interval of the anisotropy
parameters. 
In the thermodynamic limit, however,
we find that these ordering wavevectors become a continuous function 
of the anisotropy parameter, leading to a continuous set of 
incommensurate supersolids.
We emphasize that such a continuous set of supersolid orders
represent exciting discovery of novel quantum structures that have
not been seen in previous studies of lattice boson systems.

{\it Lattice Model.}---
We begin with the following hardcore boson
model on an anisotropic triangular lattice.
\begin{equation}
H_{\rm b}= \sum_{\langle i,j \rangle} \left[ -t_{ij} (b^{\dag}_i
b_j+\text{H.c.})+V_{ij} n_i n_j \right] -\mu\sum_{i} n_i,
\label{bh1}
\end{equation}
where $b_i$ denotes the boson annihilation operator at site $i$ and
$\langle ij \rangle$ runs over the nearest-neighbor sites. The
hopping $t_{ij}$ and repulsive interaction $V_{ij}$ are given by
$t_{ij}=t_1 (V_{ij}=V_1) $ and $t_{ij}=t_2 (V_{ij}=V_2)$ for the
nearest-neighbor sites along the diagonal and horizontal bonds of a
triangular lattice (if it is viewed as a square lattice with one
additional diagonal bond per each plaquette). Here we shall fix the
anisotropy parameter $\eta=t_2/t_1=V_2/V_1$. For $\eta=1$, the model
reduces to the well-known isotropic-triangular-lattice model. Such a
hardcore boson model is also equivalent to an anisotropic spin-$1/2$
XXZ model via the well-known Holstein-Primakoff mapping
\cite{trilat:ss}.

In the isotropic case, the classical limit of this model ($t_{1,2}=0$) has an
extensive ground state degeneracy and power-law density-density
correlations (or $S_z$-$S_z$ correlator in the XXZ model) at zero
temperature \cite{trilat:corr}.
The ground state degeneracy at the isotropic
point is completely lifted for $\eta<1$ and the system orders at
$\mathbf{Q}_1=(\pi,\pi)$.
On the contrary, the degeneracy is
only partially lifted for $\eta >1$. Here each diagonal chain is
ordered antiferromagnetically at $\mathbf{Q}_2=(\pi,0)$ but the
chains can be shifted with respect to each other giving rise to
$2^L$ ground states, where $L$ is the linear system size. We expect
the first (second) type of ordering in the quantum model for $\eta
\ll (\gg) 1$ and call these phases solid I(II) for future reference.
For the quantum model, when $t_{1,2}$ is turned on, it is
well-known that the system exhibits a supersolid phase at $\eta=1$
and large enough values of $V_1/t_1$ \cite{trilat:ss}. The key point
which we want to address in this paper is the fate of the supersolid
phase when $\eta \ne 1$.
\begin{figure}
\rotatebox{0}{
\includegraphics*[width=\linewidth]{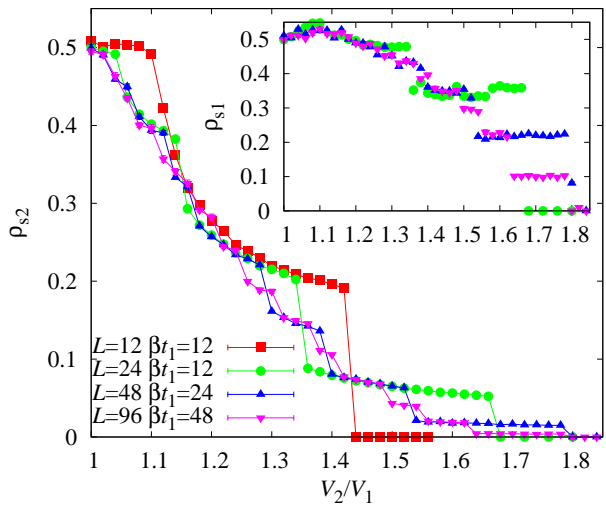}}
\caption{The superfluid density along ${\bf a_2}$ as a function
of the anisotropy parameter $\eta$ for different system sizes and
temperatures with $V_1/t_1=10$. The inset shows a similar plot for
the superfluid density along ${\bf a_1}$. Lines are guides to the eye. The number 
of steps for $L=12,24,48,96$ are $3,5,9,17$, respectively. The number of
 jumps between adjacent steps is hence $2,4,8,16$, respectively.}
\label{fig1}
\end{figure}

{\it Quantum Monte Carlo.}---
To address this problem, we perform QMC simulations using a
plaquette generalization \cite{louis:gros} of the Stochastic Series
expansion (SSE) algorithm \cite{sse}, where the elementary lattice unit is a
triangle; this results in improved efficiency for large values of
$V_1/t_1$. We measure the superfluid density along the diagonal
(${\bf a_2}$) and horizontal (${\bf a_1}$) lattice directions by
measuring the corresponding winding numbers $W^2_{\mathbf{a}_i}$
\cite{windingnumber}: $\rho_{s1(2)}= W^2_{\mathbf{a}_{1(2)}}/\beta t_{1}$,
where $\beta$ is the inverse temperature.
We also measure the equal time density-density correlator $S({\mathbf
q})/N = \langle\rho^{\dagger}_{{\mathbf q}\tau} \rho_{{\mathbf
q}\tau} \rangle$, where $\rho_{{\mathbf
q}\tau}=(1/N)\sum_i\rho_{i\tau}\exp(i{\mathbf q}\cdot {\mathbf
r_i})$ and $\rho_{i\tau}$ is the boson density at site $i$ and
imaginary time $\tau$.

\begin{figure}
\rotatebox{0}{
\includegraphics*[width=\linewidth]{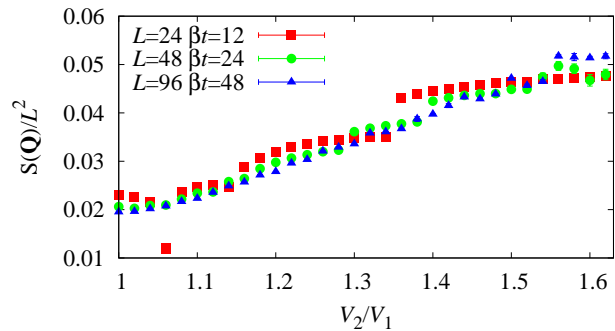}}
\caption{The equal time structure factor $S({\bf Q}(\eta))$ at
the ordering wavevector ${\bf Q}(\eta)$ as a function of $\eta$ for
different system sizes and temperatures. All other parameters are
the same as in Fig.\ \ref{fig1}. Judging from the size dependence
of the data, the structure factor is clearly finite
in the thermodynamic limit.
}\label{fig2}
\end{figure}

We begin with the case $\eta \ge 1$. A plot of the superfluid density
$\rho_{s2}$ along the $\mathbf{a}_2$ lattice direction is shown in Fig.\
\ref{fig1} as function of $\eta$ for different system sizes and
temperatures. Notice that it exhibits a staircase structure with the number of
steps proportional to the system size.  Also, when the system size is doubled,
the number of jumps between adjacent
steps doubles and the gap between them
decreases by half. The corresponding plot for
$\rho_{s1}$ is shown in the inset of Fig.\ \ref{fig1}. As can be
clearly seen from both plots, the superfluid density
undergoes several discontinuous jumps before reaching zero at $\eta
\simeq 1.8$. Also, as shown in Fig.\ \ref{fig2}, the
density-density correlator (in the thermodynamic limit) in each
segment of $\eta$ is finite at some ordering
wavevector $Q(\eta)$ in the corresponding parameter range.
Further, the ordering wavevector ${\bf Q}(\eta)$ is
a constant along the ``plateau" and changes discontinuously
upon entering the next phase.
Thus, for these finite size systems, the ``plateaus" for $1 \le \eta \le 1.8$
correspond to distinct supersolid phases with
sharp transitions between them.
%

\begin{figure}
\rotatebox{0}{
\includegraphics*[width=\linewidth]{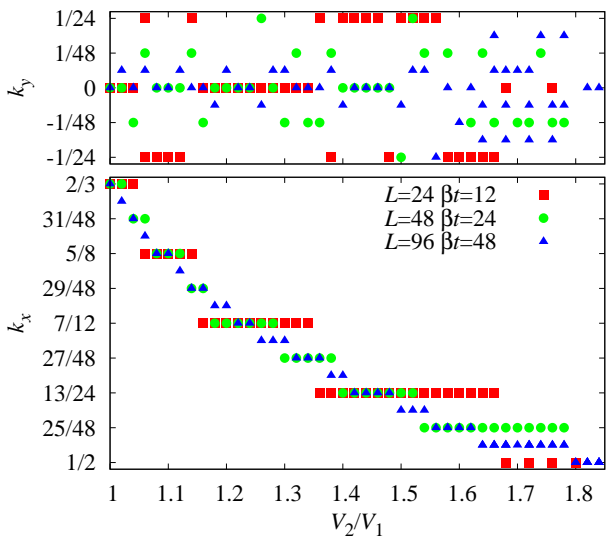}}
\caption{The ordering wavevectors ${\bf Q}(\eta)=(k_x,k_y)$
as a function of $\eta$ for different system sizes and temperatures.
All the parameters are the same as in Fig.\ \ref{fig1}. Note that for odd
$k_x L$, $k_y$ cannot be zero and hence is off the axis.
The data for $L=96$ for $\eta > 1.6$ are not fully equilibrated,
leading to the noisy behavior.}
\label{fig3}
\end{figure}

Next we address the anisotropy dependence of the ordering
wavevectors. The evolution of ${\bf Q}=(k_x,k_y)$ as a function of
$\eta$ is shown in Fig.~\ref{fig3}. Here we have chosen
$(k_x,k_y/\sqrt{3}) = (n_1 {\bf b_1} + n_2{\bf b_2})/2 \pi L$, where
$\mathbf{b}_1=2\pi(1,1/\sqrt{3}),\mathbf{b}_2=2\pi(0,2/\sqrt{3})$
are the reciprocal lattice vectors and $n_1,n_2$ are integers.
Comparing Fig.\ \ref{fig3} and \ref{fig1} we find that the system
locks at rational wavevectors that are commensurate with the
lattice in the ``plateau" regions. The ordering wavevector goes from
$\mathbf{Q}_0=(4\pi/3,0)$ at the isotropic point ($\eta=1$) to
$\mathbf{Q}_1=(\pi,0)$ at the transition point to the solid II phase
($\eta \simeq 1.8$), picking all possible commensurate values in
between. The discontinuous jumps of ${\bf Q}(\eta)$ between these
commensurate values decrease in magnitude with increasing system size.
The nature of the phase diagram for $\eta <1$,
with $0.84 \le \eta \le 1$, turns out to be qualitatively similar.
We again find a series of supersolid phase with the ordering
wavevector pinned to commensurate values along the ``plateaus", before
the system reaches the Mott phase solid I at $\eta=0.84$. The only
difference comes from the fact that for $\eta <1$, $k_xL$ is always
even, so that $k_y$ remains pinned to zero throughout the phase
diagram.

{\it Variational Monte Carlo.}---
We now supplement the QMC results with the VMC
studies of the global phase diagram. Following Ref.\ \cite{sorella1}, we
use a variational wave function:
\begin{equation}\label{wavefunction}
|\Psi \rangle = e^{-\frac{1}{2} \sum_{i,j} v_{i,j} n_i n_j } |\Phi_0
\rangle,
\end{equation}
where $|\Phi_0 \rangle = (b_{k=0}^\dag)^N|0 \rangle$ is the
non-interacting superfluid wave function and $N$ is the total number
of bosons. The components of the Jastrow potential, $v_{i,j}=
v(|R_i-R_j|)$ are independently optimized to take into account the
correlations between particles at different sites. The variational
ground state energy decreases with larger number of $v_{i,j}$. For the
present study, we have incorporated $15$ $v_{i,j}$ parameters to
make our result qualitatively and semi-quantitatively consistent
with the QMC results.
Standard Metropolis algorithm is used to
calculate the variational energy. The method of statistical reconfiguration
by Sorella \cite{sorella1} is employed to obtain the optimized parameters.
With the optimized wave function, we compute the superfluid
density $\rho({\bf k}) = \sum_{i,j} e^{i(R_i- R_j)\cdot {\bf k}}\langle
b^{\dagger}_{i}b^{\vphantom{\dagger}}_{j}\rangle$, the density-density
correlator $S({\bf q})$, and obtain the ground state phase diagram from
these quantities\cite{vR}.
\begin{figure}
\rotatebox{0}{
\hspace{-0.5cm}
\includegraphics*[width=1.05\linewidth]{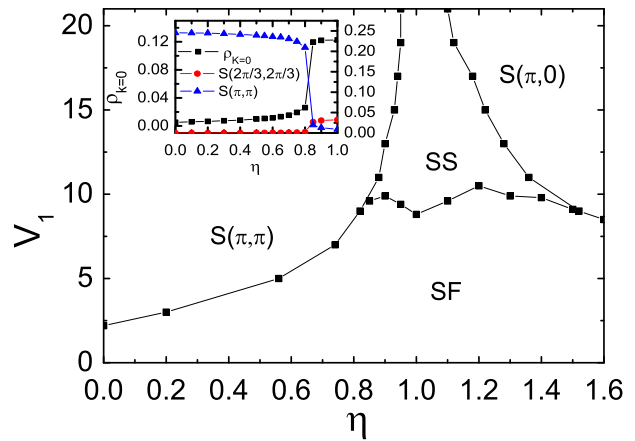}}
\caption{Ground state phase diagram obtained by
VMC as a function of $\eta$ and $V_1/t_1$ for
$L=24$. The supersolid phase exists for $V_1 \ge 10 t_1$
and between $0.8 \le \eta \le 1.5$. The inset shows the superfluid
order parameter and the density-density correlators $S(\pi,\pi)$ and
$S(2 \pi/3,2\pi/3)$ as a function of $\eta$ for $V_1/t_1=9$. The
transition from the Mott solid I to supersolid phase occurs at $\eta
\simeq 0.8$.} \label{fig4}
\end{figure}

The VMC results are summarized in Fig.\
\ref{fig4} where the ground state phase diagram is
shown as a function of $\eta$ and $V_1/t_1$ for $L=24$. Notice that
a supersolid phase exists in the range $0.8 \le \eta \le 1.5$ for
$V_1/t_1 \simeq 8$, which is qualitatively consistent with the QMC
results. We have found that the upper limit of the phase boundary
$\eta_u \simeq 1.5$ depends on the system size, and progresses
towards larger values with increasing $L$. The lower phase boundary
$\eta_l \simeq 0.8$ is virtually independent of the system size. For
$\eta \le \eta_l$($\ge \eta_u$), the system enters the Mott phase
I(II) provided $V_1/t_1 \ge 8$. For weaker interactions, the
superfluid phase prevails for all $\eta$. The inset of Fig.\
\ref{fig4} shows the transition from the Mott I to the supersolid
phase at $\eta \simeq 0.8$ for $L=12$ and $V_1/t_1=9$. Note that the
density-density correlator $S(2 \pi/3,2 \pi/3)$ and the superfluid
order parameter $\rho({\bf k=0})$ rise sharply while $S(\pi,\pi)$
drops to zero around this point, signifying a transition from the
Mott I to the supersolid phase.


\begin{figure}
\rotatebox{0}{
\hspace{-0.5cm}
\includegraphics*[width=0.85\linewidth]{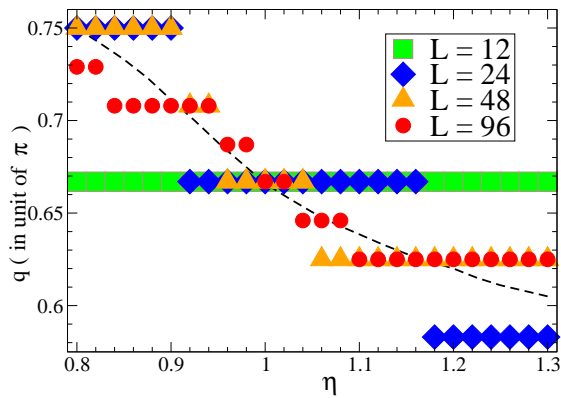}}
\caption{(Color online) Plot of $q$ (ordering wavevector is given by
$\vec{Q}=(q,q)$) as a function of $\eta$ for several finite system sizes 
(colored symbols)
and in the thermodynamic limit (dashed line) 
as obtained from Schwinger boson mean-field analysis.} \label{fig5}
\end{figure}

{\it Schwinger Boson analysis.}--- Next, to obtain an analytical
understanding of the $\eta$ dependence of the ordering wavevector,
and to determine the fate of the ordering wavevector plateaus in the
thermodynamic limit, we carry out a mean-field Schwinger boson
analysis of Eq.\ \ref{bh1}. To this end, we start from the  XXZ spin
model description of Eq.\ \ref{bh1} \cite{trilat:ss} and rewrite
these spins in terms Schwinger bosons: $S_i^{+}= a_i^{\dagger} c_i$
and $S_i^z= (a_i^{\dagger}a_i -c_i^{\dagger} c_i)/2$. Following
standard procedure \cite{sch1}, we decouple the resultant
Hamiltonian using the mean-field order parameters $A_{\alpha} =
\langle a_{i}^{\dagger}a_{i+\alpha} +c_i^{\dagger} c_{i + \alpha}
\rangle$ and $B_{\alpha} = \langle a_{i} c_{i + \alpha}-c_i
a_{i+\alpha}\rangle$, where $\alpha=x,y$ for the horizontal/vertical
and $\alpha=z$ for the diagonal bonds emanating from site $i$, to
get the Schwinger boson mean-field free energy
\begin{eqnarray}
f_{\rm MF} &=&  \frac{A_x^2 + A_y^2 + A_z^2}{\eta (1+V_1/t_1)} -
\frac{B_x^2 + B_y^2 + B_z^2}{\eta(1+V_1/t_1)} \nonumber\\
&& - (S+1/2)\lambda -\frac{1}{N} \sum_{\bf k} A_{\bf k} +
\frac{1}{2N} \sum_{\bf k} \omega_{\bf k}. \label{sbmf}
\end{eqnarray}
Here $A_{\bf k} = A_x \cos(k_x) +A_y \cos(k_y) + A_z \cos(k_x+k_y)$,
$B_{\bf k} = B_x \sin(k_x) +B_y \sin(k_y) + B_z \sin(k_x+k_y)$,
$\omega_{\bf k} = \sqrt{ |\lambda -A_{\bf k}|^2-B_{\bf k}^2}$ is the
spinon dispersion, and the parameter $\lambda$ is used to enforce
the constraint of $2S= a_i^{\dagger} a_i +c_i^{\dagger} c_i$ at the
mean-field level.

We obtain the values of mean-field variables at ground state
by solving the saddle-point equations $\frac{\partial f_{MF}}{\partial A_\alpha}
=  \frac{\partial f_{MF}}{\partial B_\alpha} = \frac{\partial f_{MF}}{\partial \lambda} = 0$.
The minima of the spinon dispersion $\omega(k)$ at $k_{min} = \pm(q/2,q/2)$
gives the spin order of the XXZ model with ordering wave-vector $\vec{Q}=
(q,q)$ as shown in Fig. 5, both at finite sizes and in the thermodynamic
limit. In the thermodynamic limit, for small $t_1/V_1 \simeq 0.04$, we
find the supersolid phase for $0.8 \le \eta \le 1.3$ with {\it
continually varying ordering wave-vector ${\bf Q}(\eta)=(q,q)$} as
shown in Fig.\ \ref{fig5}. The corresponding spinon dispersion is
gapless around both ${\bf k}=0$ (which is a signature of
superfluidity) and ${\bf k}=\pm (q/2,q/2)$ (which signifies the 
long-ranged solid order). For finite-size systems, we find that 
the spinon dispersion at ${\bf k}=\pm (q/2,q/2)$ acquires a gap 
which decreases with increasing system size and vanishes in the thermodynamic limit. Such
a gap of the spinon dispersion leads to the staircase behavior of
the ordering wave-vector as shown in Fig.\ \ref{fig5}. Our
analytical Schwinger boson results are in qualitative agreement with
both QMC and VMC results for finite system sizes and we therefore
expect it to predict the correct behavior of the ordering
wave-vector in the thermodynamic limit.

To conclude, we found that the hard-core boson system with competing
interactions on anisotropic triangular lattices is locked to a series
of commensurate supersolid phases for finite size systems, separated
by series of jumps. This behavior is expected to occur, for example,
in cold atom (bosons) systems on finite size optical lattices.
In the thermodynamic limit, however, the ordering
wavevector ${\bf Q}(\eta)$ becomes a continuous function of $\eta$,
leading to a smooth crossover between a continuous set of novel
incommensurate supersolids phases.

This work was supported by the NSERC, CIFAR, CRC, and
KRF-2005-070-C00044 (SV and YBK);
NSC 95-2112-M-009-049-MY3 (JJW and CHC)
and NSC 96-2112-M-029-003-MY3 (HCC and YCC) via
the MOE ATU Program of Taiwan. Some of the numerical
works was supported by the
National Center of High Performance Calculation
and the NCTS of Taiwan.
We also thank Roderich Moessner for helpful discussions.


\end{document}